\input harvmac

\Title{\vbox{\baselineskip12pt
\hbox{BCUNY-HEP/99-01}
\hbox{hep-th/9910122}}}
{\vbox{\centerline{Self-Gravitating Strings and String/Black Hole 
Correspondence}}}

\baselineskip=12pt
\centerline {Ramzi R. Khuri\footnote{$^*$}{e-mail: 
khuri@gursey.baruch.cuny.edu. Supported by NSF Grant 9900773 
and by PSC-CUNY Award 669663.}} 
\medskip
\centerline{\sl Department of Natural Sciences, Baruch College, CUNY}
\centerline{\sl 17 Lexington Avenue, New York, NY 10010
\footnote{$^\dagger$}{Permanent address.}}
\medskip
\centerline{\sl Graduate School and University Center, CUNY}
\centerline{\sl 365 5th Avenue, New York, NY 10036}
\medskip
\centerline{\sl Center for Advanced Mathematical Sciences}
\centerline{\sl American University of Beirut, Beirut, Lebanon
\footnote{$^{**}$}{Associate member.}}

\bigskip
\centerline{\bf Abstract}
\medskip
\baselineskip = 20pt 
In a recent essay, we discussed the possibility of using polymer sizing 
to model the collapse of a single, long excited string to a black hole. 
In this letter, we apply this idea to bring further support to 
string/black hole correspondence. In particular, we reproduce Horowitz 
and Polchinki's results for self-gravitating fundamental strings and 
speculate on the nature of the quantum degrees of freedom of black holes 
in string theory.

\Date{October, 1999}

\def\({\left (}
\def\){\right )}
\def\[{\left [}
\def\]{\right ]}

\lref\thorn{See C. B. Thorn, hep-th/9607204 and references therein; see also
O. Bergman and C. B. Thorn, Nucl. Phys. {\bf B502} (1997) 309.}

\lref\bekhawk {J. Bekenstein, Lett. Nuov. Cimento {\bf 4} (1972) 737;
Phys. Rev. {\bf D7} (1973) 2333; Phys. Rev. {\bf D9} (1974) 3292;
S. W. Hawking, Nature {\bf 248} (1974) 30; Comm. Math. 
Phys. {\bf 43} (1975) 199.}
 
\lref\GSW {M. B. Green, J. H. Schwarz and E. Witten,
{\it Superstring Theory}, Cambridge University Press, Cambridge (1987).}

\lref\prep{See M. J. Duff, R. R. Khuri and J. X. Lu, Phys. Rep.
{\bf B259} (1995) 213, M. Cvetic and D. Youm, Phys.Rev. {\bf D54}
(1996) 2612, M. Cvetic and A. A. Tseytlin, Nucl. Phys.
{\bf B477} (1996) 499 and references therein.} 

\lref\rahm{J. Rahmfeld, Phys. Lett. {\bf B372 } (1996) 198.}

\lref\pol{J. Polchinski hep-th/9611050 and references therein.}

\lref\stva {A. Strominger and C. Vafa, Phys. Lett. {\bf B379}
(1996) 99; J. Maldacena, hep-th/9607235 and references 
therein; K. Sfetsos and K. Skenderis, hep-th/9711138;
R. Arguiro. F. Englert and L. Houart, hep-th/9801053.}

\lref\corr{L. Susskind, hep-th/9309145.}

\lref\horpol{G. T. Horowitz and J. Polchinski, 
Phys. Rev. {\bf D55} (1997) 6189.}

\lref\self {G. T. Horowitz and J. Polchinski, Phys. Rev. {\bf D57}
(1998) 2557. See also S. Kalyana Rama, Phys. Lett. {\bf B424} (1998) 39.}

\lref\random{P. Salomonson and B. S. Skagerstam, Nucl. Phys.
{\bf B268} (1986) 349; Physica {\bf A158} (1989) 499;
D. Mitchell and N. Turok, Phys. Rev. Lett. {\bf 58} (1987) 1577;
Nucl. Phys. {\bf B294} (1987) 1138.}

\lref\polytext{See M. Doi and S. F. Edwards, {\it The Theory of 
Polymer Dynamics}, Clarendon Press, Oxford (1986) and references
therein.}

\lref\poly{S. F. Edwards and M. Muthukumar, J. Chem. Phys. {\bf 89}
(1988) 2435; S. F. Edwards and Y. Chen, J. Phys. {\bf A21}
(1988) 2963.}

\lref\callone{D. J. E. Callaway, Phys. Rev. {\bf E53} (1996) 3738.}

\lref\calltwo{D. J. E. Callaway, PROTEINS: Structure, Function
and Genetics {\bf 20} (1994) 124.}

\lref\apostol{T. M. Apostol, {\it Introduction to Analytic Number
Theory}, Springer Verlag (1976).}

\lref\khuri{R. R. Khuri, Mod. Phys. Lett. {\bf A13} (1998) 1407.}

\lref\damven{T. Damour and G. Veneziano, hep-th/9907030.}



Susskind's postulate of string-black hole correspondence \corr\ 
represents the most promising attempt to understand the physical 
basis of quantum gravity from string theory. In its most basic form,
it suggests a one to one correspondence between fundamental string
states and black hole states.

This correspondence was clarified further in \horpol, where it was
argued that by adiabatically increasing the string coupling constant
$g$ a string state would turn into a black hole at $g=g_c \sim N^{-1/4}$,
where $N >> 1$ is the level number of a long, fundamental
 string state. The string coupling
takes this critical value precisely when the Schwarzschild radius,
$R_S$, becomes
of the order of the string scale, $l_s$. For 
couplings below $g_c$, the picture 
of a string state prevails, while for coupling above $g_c$ the black hole
picture prevails. At this critical transition point, the string entropy
makes a smooth transition into the Bekenstein-Hawking black hole area
entropy law \bekhawk.

The authors of \horpol\ took this correspondence a step further in
\self, using a thermal scalar field theory formalism to study the 
size of the string state as it collapses from its initial random walk
\random\ form into a black hole. It was then argued in \khuri\ that 
these results should also arise using the methods of polymer 
physics\foot{The concepts of the ``string bit" and the ``polymer string"
were first discovered by Thorn \thorn.}.
More recently, the results of \horpol\ were also reproduced in \damven, 
and where some of the physical issues relating to the collapsing string 
were further clarified.

In this letter, we reproduce the results of \self\ using polymer methods,
making precise the scaling arguments of \khuri\ and thereby giving further
support to string-black hole correspondence. We also discuss the implications
of these results to the understanding of the underlying degrees of 
freedom of quantum black holes in string theory.

Following \self, we consider a long, self-gravitating string at level $N$
and  adiabatically increase the coupling $g$ until the string collapses 
into a black hole. As noted in \self, the string size at zero coupling 
(the free string) is initially given by $R_{RW} \sim N^{1/4} l_s$, where 
$l_s$ is the string scale. The letters ``RW" denote ``Random Walk", as the 
free string represents a random walk \random\ with $n=N^{1/2}$ steps (or 
string ``bits" \thorn). 
The total length of the string is given by $L=nl_s$.

As argued in \khuri, this configuration may be represented as a 
random walk polymer chain with self-interaction. There are $n$ steps,
each of length $l$, with $\vec r_i$ representing the position of the 
chain after the $ith$ step. Gravitational self-interactions start to become 
significant once $g \sim g_0 \sim n^{(d-6)/4}$. This can be seen immediately
by noting that the attractive potential
\eqn\potential{U=\sum_{i\neq j} {g^2\over |\vec r_i - \vec r_j|^{d-2}}
\sim {n^2 g^2\over n^{(d-2)/2} l_s^{d-2}} \sim g^2 n^{-(d-6)/2} 
(l_s^{2-d})}
becomes non-neglible in string units only when $g$ approaches $g_0$. Here we 
have used the fact 
that the mean square average of the distance between steps (the radius of
gyration \polytext) is proportional to the random walk size, $R_{RW}\sim 
n^{1/2} l_s$. We then replaced accordingly in the summation, up to a factor of the order unity (the precise numerical coefficient is not necessary for 
our purposes).

As in \khuri, we follow the calculation of \poly\ but instead of a 
self-avoiding delta-function interaction, we consider a self-attracting
potential $U$ as above. For simplicity, we restrict ourselves initially
to the case of $D=4$ spacetime dimensions (or $d=3$ space dimensions).

The system is described by the generalized Hamiltonian
\eqn\hamilt{\beta H ={3\over 2l}\int_0^L ds \left({\partial \vec R(s)\over 
\partial s}\right)^2 + g^2l \int_0^L \int_0^L ds ds'{1\over 
|\vec R(s) -\vec R(s')|},}
where $\vec R(s)$ is the position vector of the chain at arc-length
$s$ ($0 \leq s \leq L$). Following \poly, we use the Feynman variational
procedure for the free energy of the chain. The trial Hamiltonian is
given by
\eqn\trial{\beta H_0={3\over 2l_s} \int_0^L ds \left({\partial \vec 
R(s)\over \partial s}\right)^2 + Q,}
where 
\eqn\qqq{Q={q^2\over 6l_s} \int_0^L ds \vec R(s)^2,}
where $q$ is the variational parameter.
The free energy of the chain is given by
\eqn\freeone{\exp(-\beta F) = \int D\vec R \exp(-\beta H_0 + X + Q),}
where
\eqn\ex{X={g^2\over l_s} \int_0^L \int_0^L ds ds' {ds ds'\over 
|\vec R(s) -\vec R(s')|},}
and where $q$ is a variational parameter. $F$ is to be extremized with 
respect to $q$. Straightforward calculations similar to those 
of \poly\ yield the free energy 
\eqn\freetwo{\beta F = \beta H_0 - Q - X = {qL\over 4} -\epsilon 
{g^2 L^2\over l} \left({q\over l_s}\right)^{1/2},}
where $\epsilon$ is a number of the order unity.
It is also straightforward to show that the size of the polymer,
the average mean square end-to-end distance of the chain, is given by
\eqn\sizeone{R^2 = {l_s\over q_0} (1-\exp{(-q_0L)}),}
where $q_0$ is the value of $q$ obtained by varying \freetwo\
with respect to $q$ and is determined up to a factor of $O(1)$. The 
variation of \freetwo\ yields $q_0 L \sim g^4 n^3$, which implies
\eqn\sizetwo{R^2 \simeq {l_s^2\over g^4 n^3} \left(1 - \exp{(-g^4n^3)}\right).}
This finding makes precise the arguments of \khuri, and agrees with 
the results of \self. In particular, note that for $g << g_0$,
$R^2 \sim nl_s^2 $, which is the random walk/free string result, while for
$g_0 < g < g_c$, $R \sim l_s/(g^2n)$, which agrees with the calculation
of \self.

The above results can be generalized in a straightforward manner to the
case of arbitrary spacetime dimension $D=d+1$ (note, however, that the 
calculations are not reliable for $d > 4$). In this case, the free 
energy is given by 
\eqn\freegen{\beta F = = {qL\over 4} -\epsilon 
{g^2 L^2\over l_s^{4-d}} \left({q\over l}\right)^{{1\over 2}(d-2)},}
where again $\epsilon$ is a number of the order unity.
The size of the polymer is again given by \sizeone. The variation for
general $d\neq 4$ yields $q_0 L \sim (g^4 n^{6-d})^{1/(4-d)}$, which again
predicts a transition from the random walk at $g_0 \sim n^{(d-6)/4}$, although the nature of
the transition is somewhat more obscure \refs{\self,\damven}.

The special case of $d=4$ ($D=5$) deserves some attention. Here
$F \sim qL(1-\epsilon g^2n)$, so that for $g < g_c \sim n^{-1/2}$,
$F$ is minimized at $q=0$, corresponding to the random walk, while
for $g > g_c$, $F$ is minimized for $q\to \infty$, so that 
$R^2\to 0$, corresponding to a rapid collapse at the transitional 
coupling, which coincides in this case with the critical coupling:
$g_0 = g_c$. This is again in agreement with the results of \self\ and 
\damven. Note that the five-dimensional case also corresponds to
the expected behaviour for a string state winding around the fifth dimension,
and whose collapse results in a charged black hole.

Further agreement with the results of \self\ can be seen by considering
the temperature.
To leading order, $T=T_H=1/\beta =1/l_s$, where $T_H$ is the Hagedorn
temperature, so we 
will compute the leading order correction.
Again, for simplicity, let us focus on the case of $D=4$ ($d=3$).
Then at $q=q_0$, $\beta F = \beta (E-TS) \sim g^4 n^3$. 
Using $E = M \sim n/l_s$, we obtain, to leading order,
$TS \sim n/l_s(1-\epsilon g^4n^2)$. Since $S\sim n$ throughout the
adiabatic collapse, this gives $T \sim (1-\epsilon g^4n^2)/l_s$. 
This again reproduces the results of \self, namely that 
$\beta - \beta _H \sim g^4 n^2 l_s \sim g^4 M^2 l_s^3 \sim \alpha'^2g^4M^2/\beta_H$.

Let us now see what our polymer-type picture says about the nature of the
degrees of freedom in the transition between the string and the black hole
pictures. Consider the entropy $S$ of the polymer string in $D=4$
(the arguments are similar in other dimensions in which the above
transition makes sense).
At zero coupling, $S_0 \sim n$, the number of steps. This can be seen 
in the usual manner for a random walk by associating to each step a
a small whole number of independent possible directions (related to the 
number of space dimensions) $p$. The number of possible configurations 
gives the degeneracy $d \sim p^n$, so that the entropy $S = \ln d \sim n$.

In the intermediate range $g_0 < g < g_c$, the adiabatic increase of the
coupling preserves the essential degrees of freedom associated with the
string bits. Of course one no longer has a random walk, but up to a
factor of order unity, the string bits retain most of their degrees of
freedom.

For a black hole whose mass is equal
to the excited string state up to a factor of $O(1)$, 
$S_{BH}=A/4G \sim R_S^2/l_P^2$, where $l_P=g l_s$ is the Planck scale and 
$R_S \sim GM \sim l_P^2 M \sim l_P^2n/l_s \sim g^2n l_s$ 
is the Schwarzschild radius. We may then rewrite the BH entropy
as
\eqn\bhent{S_{BH} \sim {n^2 l_P^2\over l_s^2}\sim n^2 g^2.}
At the critical coupling $g_c \sim n^{-1/2}$, the entropy makes a smooth transition to the Bekenstein-Hawking area law form: $S_{BH} \sim n \sim S_0$. This makes sense, since we don't expect any violent changes in the entropy
in the process of adiabatically increasing the coupling. It is interesting,
however, that this entropy still represents a degeneracy of a polymer 
system with $n$ steps. This can be seen as follows: at $g=g_c$, the size
of the collapsed polymer string is given by $R \sim R_S \sim l_s$, or the
size of one string bit, or one step. In order for $n$ steps of size $l_s$
to fit into a sphere of radius $R_S$, the number of possible positions to 
which each step can go must remain a small whole number, $p'$. Hence the 
degeneracy again has the random walk form $d \sim p'^n$.
Of course, the polymer is no longer a random walk, but the degrees of 
freedom still remain essentially intact. 

What happens in the black hole picture once the transition is complete?
In this case, the area of the black hole is given by 
$A\sim R_S^2 \sim l_s^2 \sim (1/g^2) l_P^2 \sim n l_P^2$. So the horizon
can be divided into $n$ ``pixels" each of area $l_P^2$. Once the horizon forms,
the degrees of freedom associated with it represent independent quantum
states\foot{The independence of the states is a result of the causally 
disconnected nature of the points on the horizon.}. Again, only a small whole number of possible states, 
$q$, is associated with each pixel, so that the total degeneracy is given
by $d_{BH} \sim q^n$, with the entropy given by 
$S = \ln d \sim n \sim S_{BH}$. 
So essentially the random walk degrees of freedom turn into horizon
surface degrees of freedom at the critical transition point. Another way of saying this is that the string bits project their information onto the horizon.

What happens if we continue increasing the coupling beyond the critical
coupling $g_c$ and into the black hole domain? Then the surface degrees
of freedom should continue to account for the area law entropy. This
can be seen in two ways: in the first way, we fix the mass while 
increasing $g$. This would necessarily no longer be an adiabatic change,
and the entropy would have to change. In this case, 
$R_S \sim g^2 n l_s > l_s$ for $g > g_c = n^{-1/2}$, and the area 
$A \sim R^2 \sim g^4 n^2 \l_s^2 \sim g^2 n^2 l_P^2$. The number of 
pixels is now given by $k=A/l_P^2 = g^2 n^2$, which is equal
to Bekenstein-Hawking entropy, $S_{BH}$ from \bhent.

The other way of extending beyond $g_c$ is to continue adiabatically. 
In this case the entropy remains equal to the original random walk number
$n$, but the mass changes so that $S \sim A/G \sim R_S^2/G \sim GM^2 \sim n$.
This implies that $M^2 \sim n/l_P^2 \sim n/(g^2l_s^2)$, so that 
$M\sim \sqrt{n}/(gl_s)$. It follows that 
$R_S \sim GM \sim S/M \sim ngl_s/\sqrt{n} = \sqrt{n} gl_s = \sqrt{n} l_P$,
and the horizon area is $A \sim R_S^2 \sim n \l_P^2$, representing 
$n$ pixels and a quantum entropy of $n$. 

The above results are in full agreement with expectations of the holographic
principle, namely that the degrees of freedom of the black hole can
be recovered through projection of the states onto the horizon surface.
In conclusion, the above results not only independently confirm those of 
\refs{\self,\damven}, but also suggest that the underlying degrees of 
freedom of quantum black holes in string theory remain associated 
with the original, stringy degrees of freedom. This further strengthens
the string/black hole correspondence conjecture \corr\ by implying that
in the transition to the strong-coupling limit, it is possible that the
quantum string states somehow retain far more of their nature from the
perturbative picture than might have been supposed. This idea, of course, 
requires further verification and more precise calculations, but 
nevertheless points to the robustness of the string/black hole 
correspondence principle.

\listrefs
\end